\documentclass[runningheads]{llncs}
\usepackage{graphicx}
\usepackage[utf8]{inputenc}
\usepackage{booktabs} 
\usepackage{amsmath}
\usepackage{amsfonts}
\usepackage{todonotes}
\usepackage{subfig}

\newcommand{\paraHd}[1] {\noindent\textbf{#1.}} 
\newcommand{\paraHdSpace}[1] {\vspace{1.2mm}\noindent\textbf{#1.}} 

\begin{document}

\title{Replicating Relevance-Ranked Synonym Discovery in a New Language and Domain}
\titlerunning{Replicating Relevance-Ranked Synonym Discovery}
%

\author{Andrew Yates\inst{1} \and Michael Unterkalmsteiner\inst{2}}
\authorrunning{Yates and Unterkalmsteiner}
\institute{Max Planck Institute for Informatics, Saarbrücken, Germany
\email{ayates@mpi-inf.mpg.de}\\
\and
Blekinge Institute of Technology, Software Engineering Research Laboratory, 
Karlskrona, Sweden\\
\email{michael.unterkalmsteiner@bth.se}}

\maketitle              
\begin{abstract}
  Domain-specific synonyms occur in many specialized search tasks, such as when searching medical documents,
  legal documents, and software engineering artifacts. We replicate prior work on ranking domain-specific
  synonyms in the consumer health domain by applying the approach to a new language and domain:
  identifying Swedish language synonyms in the building construction domain.
  We chose this setting because identifying synonyms in this domain is helpful for
  downstream systems, where different users may query for documents (e.g., engineering requirements) using different terminology.
  We consider two new features inspired by the change in language and methodological advances
  since the prior work's publication. An evaluation using data from the building construction
  domain supports the finding from the prior work that synonym discovery is best approached as a
  learning to rank task in which a human editor views ranked synonym candidates in order
  to construct a domain-specific thesaurus.
  We additionally find that FastText embeddings alone provide a strong baseline,
  though they do not perform as well as the strongest learning to rank method.
  Finally, we analyze the performance of individual features and the differences in the domains.
\end{abstract}
\section{Introduction}
The vocabulary mismatch problem~\cite{furnas1987vocabulary} and its detrimental
effect on recall~\cite{carpineto2012survey} have long been recognized by the
information retrieval community. In the absence of query expansion, whether
by using pseudo relevance feedback or a lookup method,
finding relevant information is a matter of constructing a query that
expresses the user's information need using the same terms found in relevant documents.
In the case of domain-specific search tasks, such as search in medical documents~\cite{kang2017}, 
legal documents~\cite{Oard2010}, patents~\cite{zhang2015patent} and 
software engineering artifacts~\cite{lucia2007recovering,Haiduc2013}, domain-specific
synonyms complicate information retrieval as either the searcher or the 
retrieval engine need to be aware of terms that may be used interchangeably. 
These specialized information retrieval tasks can benefit from thesauri that are crafted for 
their specific domain.

We chose to replicate and generalize a paper on using learning to rank for human-assisted
synonym discovery, because we are interested in improving the synonyms
in a classification system (ontology) from the building construction domain.
While the users of 
the classification system are professionals, they are usually specialized in 
subsets of the construction business and need to coordinate with users specialized in other 
subsets of the domain. An improved set of synonyms would likely help the 
collaboration between the different parties using the classification system,
such as when searching requirements documents. 

Yates et al.~\cite{Yates14} proposed a method that produces a ranked 
list of domain-specific synonyms using a domain-specific corpus as 
input. Their learning to rank approach uses a set of features that outperformed 
previous synonym discovery methods that relied on single statistical measures:
pointwise mutual information over term 
co-occurrences~\cite{terra2003frequency} and pointwise total correlation 
between two terms and the syntactic context (dependency relations) in which the 
terms appear~\cite{hagiwara2008supervised}.

In this paper we replicate Yates et al.'s method on a different domain (building 
construction) and language (Swedish) in order to evaluate the generalizability of the approach.
We used the original implementation, 
adapting preprocessing and features to the new setting. As 
such, we perform an inferential reproduction~\cite{goodman2016does} where we draw 
similar conclusions as the original paper 
after evaluating it on a completely different dataset.
Our contributions are \textit{(1)} an evaluation of the method proposed by Yates et al. on a new language and different domain and \textit{(2)} the proposal and evaluation of new features inspired by recent methodology advances and the differences introduced by the new domain and language.

\section{Methodology} 
In this section we describe the synonym discovery task before describing
our replication of the experimental setup used by Yates et al.
We refer to the paper by Yates et al. as the original study and to this paper as the replication study.

\subsection{Problem formulation}
As in the original study, we define \textit{synonym discovery} as the task of identifying
a target term's correct synonyms from among a set of synonym candidates. Due to the difficulty
of this problem, the original study argued that it is best approached as a synonym search
task in which a domain-specific corpus is coupled with a learning to rank method in order to
help a user quickly identify a target term's synonyms. For example, in the original study,
the user might search for the target term \textit{alopecia} with the intent of identifying
domain-specific synonyms such as \textit{hair loss} and \textit{missing hair}. Such synonyms
would ideally be ranked highly, but because of the task's inherent difficulty, incorrect terms
like \textit{greying hair} and \textit{headache} are also likely to be ranked highly.
This ranked list can then be used by a human editor in order to manually build a list of
domain-specific synonyms. For example, in the original study the method was used to augment
a thesaurus mapping expert medical terms (e.g., \textit{alopecia}) to lay synonyms
often used in social media in place of the expert terms (e.g., \textit{losing hair}).

The original study proposed to re-frame the 
evaluation of synonym discovery approaches from a TOEFL\footnote{Test Of 
English as a Foreign Language} style problem (i.e., given a 
term, pick the correct synonym from $n$ candidate terms) to a ranking problem 
(i.e., given a term, evaluate how many true synonyms are ranked in the top X\% of $n$ 
candidates). This problem re-framing stems from the observation that the
TOEFL style test represents the synonym discovery problem only when there is a sufficiently 
large number of synonym candidates. However, by increasing the number of 
incorrect choices, the evaluated approaches, including their own, were not able 
to answer the TOEFL-style question in most cases. Ranking synonym candidates 
according to their probability of being a true synonym of a target terms mirrors 
the synonym discovery problem more accurately. Ultimately, the involvement of a human
editor is required to build an accurate domain-specific thesaurus.

One motivation for performing domain-specific synonym discovery is that
we would like to cater for both propositional synonyms and near-synonyms. 
Propositional synonyms refer to terms that can be used interchangeably without 
affecting the truth condition of a statement. For example, the statements 
\textit{He is a statesman} and \textit{He is a politician}, referring to the 
same person, can both be true. This type of synonymity is concerned with 
identity rather than similarity of meaning, while near-synonyms refer to terms 
with similar meaning and are context-dependent~\cite{stanojevic2009cognitive}. 
For example, in the building construction domain, the term \textit{kylelement} 
\textit{(cooling panel)} and the term \textit{förångare} \textit{(evaporator)} 
are synonymous, and are used to describe a thermal cooling object. However, in 
the maritime domain, evaporators are used off-shore to produce fresh water, 
a different function from the building construction domain. 
These subtle domain-specific differences call for an approach that takes 
context into account and allows for the inclusion of human expertise when 
selecting near-synonyms from synonym candidates. 

Formally, given a target term $w_t$ and a set of candidate terms $\mathbb{C}$, a synonym discovery
method ranks each candidate term $w_c \in \mathbb{C}$ with respect to its likelihood of being a
synonym for the target term $w_t$ in the target domain. 

\subsection{Replication design}
The objectives of this replication study are to \textit{(1)} evaluate the generalizability
of the original finding that synonym discovery is best approached as a ranking problem,
\textit{(2)} to evaluate the generalizability of the learning to rank method proposed
by Yates et al. in order to determine whether it is still the best approach in a new language and domain,
and \textit{(3)} to investigate whether the approach can be improved in our new setting by incorporating
methodological improvements that were not considered in the original work (i.e., a language-specific feature,
a contemporaneous term embedding feature, and a more sophisticated learning to rank model).
We generalize over both the experimental setting (i.e., language and domain) and over time by considering
both features specific to our new setting and new approaches that have become popular since the original work's publication.
We evaluate the original approach, the best-performing baseline in the original work, a new embedding-based
baseline, and a variant of the original approach using an improved learning to rank model on both the
TOEFL task and the relevance ranking task from the original work.

\subsection{Baselines}
We compare the learning to rank method proposed in the original study with several baselines.
We include the baseline that performed best in the original study as well two additional baselines
based on embedding similarity and the similarity of dependency relation contexts.

\paraHdSpace{PMI} The best-performing baseline in the original study was pointwise
mutual information (PMI) calculated over term co-occurrences in sliding windows as
proposed by Terra and Clarke \cite{terra2003frequency}. We calculate PMI over 16-term sliding windows
with the constraint that each window can only cover a single sentence.
$PMI(w_c, w_t)$ can then be used as the ranking function for obtaining a ranked list of synonyms
for the target term $w_c$.

\paraHd{EmbeddingSim} Word embeddings have become common since the original study's publication
and are often used in information retrieval and natural language processing tasks. These methods
are trained in an unsupervised manner on a large corpus to create dense word representations that
encode some of the words' properties. The FastText \cite{bojanowski2017enriching}, word2vec \cite{word2vec},
and GloVe \cite{pennington2014glove} methods are often used.
Similar to the PMI method, word embeddings capture distributional similarity, and work has shown
that much of their improvements over PMI come from the training setup used rather than from the
underlying algorithm \cite{TACL570}.
We train FastText on our corpus and rank a target term's candidates based on the cosine similarity
between the term embeddings for $w_t$ and $w_c$.
The incorporation of this baseline is an example of ``generalization over time'' since this method
is clearly relevant to the synonym discovery task, but it was not available at the time of the original
study's publication.

\paraHd{LinSim} Lin's similarity measure \cite{lin1998automatic} was originally proposed as a method for
identifying synonyms and other related words. LinSim was used as a feature in the original study, but
not as a separate baseline. This measure is similar to Hagiwara's methods \cite{hagiwara2008supervised}
from the original study in that it considers pointwise mutual information over term contexts defined by
dependency relations. It has the advantage of being less computationally expensive to compute on large
corpora, however, so in this study we use it in place of Hagiwara's supervised and unsupervised methods.

\subsection{Supervised Approaches}\label{sub:approach}
We evaluate two supervised learning to rank approaches on our dataset:
a logistic regression as proposed in the original study, which is a pointwise method that has been used
for learning to rank in other contexts \cite{Zeng:2004:LCW:1008992.1009030},
and LambdaMART \cite{burges2010ranknet}, a pairwise method.
Given a target term $w_t$ and a candidate term $w_c$, we compute the following features for use
with both supervised methods:

\begin{itemize}
\item \texttt{Windows}: the number of windows containing both $w_t$ and $w_c$, normalized by the smaller of the two counts. With the Wikipedia and Trafikverket corpora, a window is defined as a sequence of up to 16 terms appearing in a single sentence. With the Web corpus, a window is defined as a sequence of up to 16 terms appearing in a HTML element.
  Let $count_{win}(x)$ be the number of windows containing the term $x$ and $count_{win}(x, y)$ be the number of windows containing both terms $x$ and $y$. This feature is then calculated as
  $$ Windows(w_t, w_c) = \frac{count_{win}(w_t, w_c)}{min(count_{win}(w_t), count_{win}(w_c))} $$
\item \texttt{LevDist}: the Levenshtein distance (i.e., edit distance) between $w_t$ and $w_c$.
\item \texttt{NGram}: the probability that the target term $w_t$ appears in a specific position in a n-gram given that the candidate term $w_c$ has also appeared in this position. As in the original work, we consider all trigrams that appear in our corpora. Let $count_{ng}(x)$ be the number of unique n-grams a term $x$ appears in and $count_{ng}(x, y)$ be the number of unique n-grams in which both terms $x$ and $y$ appear in the same position (e.g., given the trigrams \textit{rate/of/building} and \textit{rate/of/construction}, both \textit{construction} and \textit{building} appear in the same position). This feature is then calculated as
  $$ NGram(w_t, w_c) = \frac{count_{ng}(w_t, w_c)}{count_{ng}(w_c)} $$
\item \texttt{POSNGram}: the probability that the target term $w_t$ appears in a specific position in a part of speech n-gram given that the candidate term $w_c$ has also appeared in this position. As in the original work, this is equivalent to \texttt{NGram} after replacing each term with its part of speech.
\item \texttt{LinSim}: the similarity between $w_t$ and $w_c$ as computed using Lin's similarity measure \cite{lin1998automatic}. This measure requires dependency parsing, which we perform with MaltParser \cite{nivre2007maltparser}. This feature also serves as one of our baselines.
\item \texttt{RISim}: the cosine distance between the vectors for $w_t$ and $w_c$, as computed using random indexing. We compute these vectors use the SemanticVectors package\footnote{\url{https://github.com/semanticvectors/semanticvectors/}} \cite{widdows2010semantic} with its default parameters.
\item \texttt{Decompound}: the number of components shared by $w_t$ and $w_c$, normalized by the minimum number of components in either term. We use the SECOS decompounder \cite{riedl2016secos} to split each term into their components, which decompounds each term using several strategies. We always choose the decompounding strategy that results in the largest number of components.
\item \texttt{EmbeddingSim}: the cosine distance between word embeddings for $w_t$ and $w_c$. We used FastText embeddings \cite{bojanowski2017enriching} trained on our corpus. This feature also serves as one of our baselines.
\end{itemize}

As described in the baselines section, we do not consider the two features from the original work that were based on Hagiwara's definition of contexts.
The {Decompound} and {EmbeddingSim} features did not appear in the original work. We introduce the {Decompound} feature to account for the fact that many of our target terms are compound nouns; it is often the case that their synonyms share components with the target term. For example, the target term \textit{apparatskåp} (\textit{device cabinet}) shares the component \textit{skåp} (\textit{cabinet}) with its domain-specific synonym \textit{elskåp} (\textit{electrical cabinet}).
We introduce the {EmbeddingSim} feature, on the other hand, because embedding-based similarity measures based on FastText \cite{bojanowski2017enriching}, word2vec \cite{word2vec}, and GloVe \cite{pennington2014glove} have become popular alternatives to random indexing since the original work's publication. 

\section{Replication}
\subsection{Dataset}\label{sec:ds}
The original experiment focused on the medical side effect domain, with a 
corpus written in the English language. In this replication, we focus on the 
building construction business, in particular the provisioning and building of 
roads and public transportation infrastructure, and change the language to 
Swedish. We use the synonyms defined in CoClass~\cite{noauthor_coclass_2018} as 
the ground truth. 
CoClass is a hierarchical classification system, implementing ISO 
12006-2:2015~\cite{noauthor_building_2015}, that is intended to facilitate
the life-cycle management of construction projects. It is co-developed by the Swedish 
Transportation Agency (\textit{Trafikverket}) and consultancy firms. Table~\ref{tab:comp} 
provides an overview of the dataset differences between the two studies.

\begin{table}
	\caption{Comparison of the original experiment and the 
	replication}\label{tab:comp}
	\begin{tabular}{p{6cm}ll}
		\toprule
		& Original & Replication\\
		\midrule
		Domain & Medical side effects & Building construction\\
		Language & English & Swedish\\
		Terms with/without synonyms & 1,791/0 & 574/856 \\
		Average number of synonyms per term & 2.8 ($\sigma=1.4$) & 3.8 
		($\sigma=4.4$) \\
		Min/Max number of synonyms per term & 2/11 & 1/46 \\ 
		Phrases/single term proportion & 67\%/33\% & 26\%/74\% \\
		Corpus size (Number of documents) & 400,000 & 4,241,509 \\
		\bottomrule
	\end{tabular}
\end{table}

While in the original study all terms were associated with at least 2 synonyms, 
only 574 of the terms in CoClass (40\%) are associated with any synonym.
Since our goal 
is eventually to improve the classification system with newly discovered 
synonyms, we did not remove synonyms that did not appear in the initially 
constructed corpus, which was crawled from Trafikverkets' publicly
accessible document database (1,100 documents) and the Swedish Wikipedia (3,7 
million articles). Only a subset of CoClass terms were found in this corpus, however.
Therefore, we devised 
the following strategy to construct a corpus: for each term in CoClass, we 
searched the public internet for this term using the Bing Search 
API\footnote{https://azure.microsoft.com/en-us/services/cognitive-services/bing-web-search-api/},
 contributing 540,409 documents to the total corpus.
 Since each API call returns at most 50 hits, our budget was limited, and some terms in CoClass
 were common, we used a crawling strategy focused on identifying documents containing uncommon terms.
 More specifically, we restricted the number of crawled 
websites $c$ based on the number of search results $r$ for each term:
\begin{equation}
	c=
		\begin{cases}
			2500, & \text{if}\ r<=10000\\
			1000, & \text{if}\ 10000<r<100000\\
			500,  & \text{if}\ r>=100000\\
		\end{cases}
\end{equation} 

Category $c_{500}$ contained 494 search terms, while $c_{1000}$ contained 708 and 
$c_{2500}$ contained 2261 search terms. Within $c_{2500}$, 528 search terms produced no 
hits at all.
The search results demonstrate that the terminology in CoClass is
very specialized as a large amount of terms were not even found on the publicly 
accessible internet.

\subsection{Implementation Details}\label{sec:pp}
We preprocess our corpus using the efselab
toolkit\footnote{\url{https://github.com/robertostling/efselab}} \cite{ostling2018part}
to tokenize and lemmatize the input text. In the case of the Wikipedia and Trafikverket corpora,
we additionally perform sentence segmentation. On the Web corpus we use
textract\footnote{\url{https://textract.readthedocs.io}} to identify text inside of HTML elements
(e.g., between \textit{$<$p$>$} and \textit{$<$/p$>$} tags) and treat these text spans as sentences.
We use efselab to perform part-of-speech tagging and MaltParser \cite{nivre2007maltparser}
to perform dependency parsing for the features that require this information.
Our preprocessing differs slightly from the original work due to the changes in our input language
and corpus. The original work used a tokenizer based on
the Natural Language Toolkit\footnote{\url{https://www.nltk.org/}} (NLTK), the Porter stemmer,
NLTK's part-of-speech tagging, and RASP3 \cite{rasp} for dependency parsing.

The original work took advantage of large, mature domain-specific thesauri to generate
synonym candidates from the target domain. Such thesauri are not available in our language
and domain, so we were forced to consider every term that appeared in our Wikipedia or Web corpora
as a synonym candidate. We filtered these candidates by removing terms with a low term frequency,
terms that did not appear much more often in our domain-specific Trafikverket corpus than in Wikipedia,
and terms that were not tagged as a noun by our part-of-speech tagger. In particular, we required
the candidates to have a TF of at least 300, to occur at least 30 times more often in Trafikverket
than in Wikipedia, and to be tagged as a noun at least 50\% of the time. These filtering steps
reduced the total number of synonym candidates from approximately 867,000 to 26,000 (97\% reduction)
at the cost of reducing the candidates' coverage of true synonyms in CoClass by approximately 26\%
(i.e., 74\% of the synonyms remained after filtering).
This left us with 290 target terms that both appeared in our corpus and had true synonyms in our candidate list.

As in the original work, we use the logistic regression implementation from
scikit-learn\footnote{\url{http://scikit-learn.org}} and scale the features to unit variance.
We use the LambdaMART implementation from pyltr\footnote{\url{https://github.com/jma127/pyltr}}
with query subsampling set to 0.5 (i.e., 50\% of the queries used to train each base learner)
and the other parameters at their default values. For the word embedding feature, we used
FastText\footnote{\url{https://github.com/facebookresearch/fastText/}} \cite{bojanowski2017enriching}
to train 100-dimensional embeddings on our corpus using the skipgram method.
FastText's other parameters were kept at their default settings.

Our code, the CoClass ground truth, and the URLs of documents in our corpus are available online.\footnote{\url{https://github.com/andrewyates/ecir19-ranking-synonyms}}

\subsection{Experimental Setup}\label{sec:es}
We conduct two experiments in which we compare the LogReg and LambdaMART learning to rank methods
against three baselines: PMI, EmbeddingSim, and LinSim.
In the TOEFL-style evaluation, we confirm the original study's conclusion that synonym discovery
is best approached as a ranking problem. We then evaluate the methods' ability to produce useful ranked
lists of synonym candidates in the relevance ranking evaluation.

Each method receives a target term and a set of candidates as input and outputs a ranked list of
the candidates. In order to mirror the original study's evaluation, each target term is associated
with up to 1,000 incorrect candidates that are randomly sampled from the full set of candidates $\mathbb{C}$.
The supervised methods are trained with ten-fold cross validation. We create the cross validation folds
based on target terms, so each target term appears in only one fold.
We describe the metrics used by the two evaluations in their respective sections.

\begin{figure}[!ht]
	\centering
	\subfloat[TOEFL evaluation (accuracy)]{\label{subfig:acc}
		\includegraphics[width=0.48\textwidth]{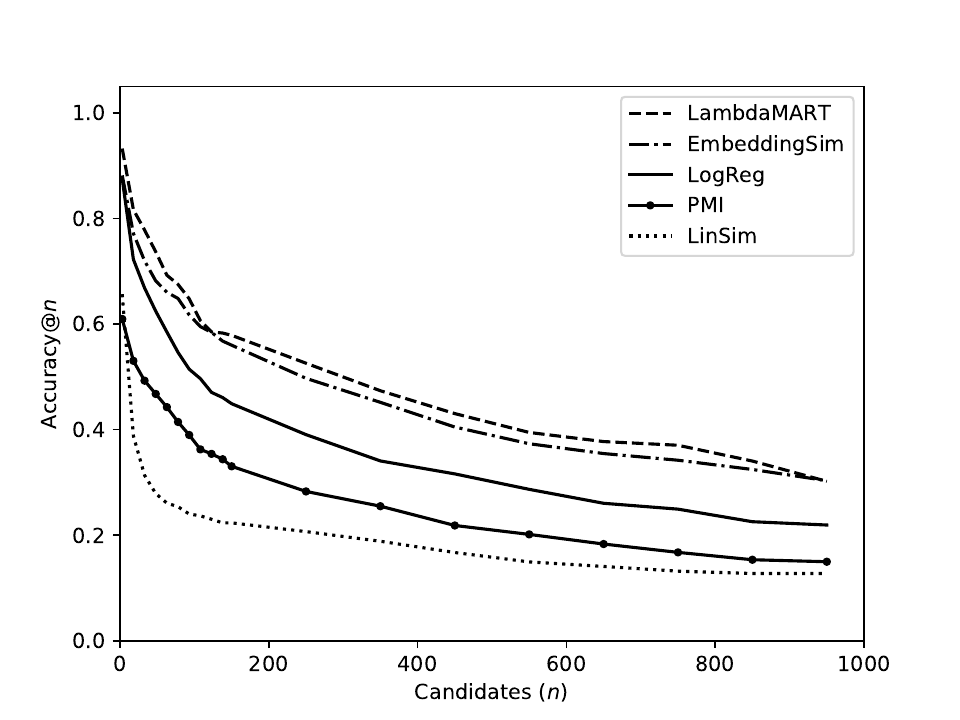}}
	\hspace{0.1cm}
	\subfloat[TOEFL evaluation (MRR)]{\label{subfig:mrr}
		\includegraphics[width=0.48\textwidth]{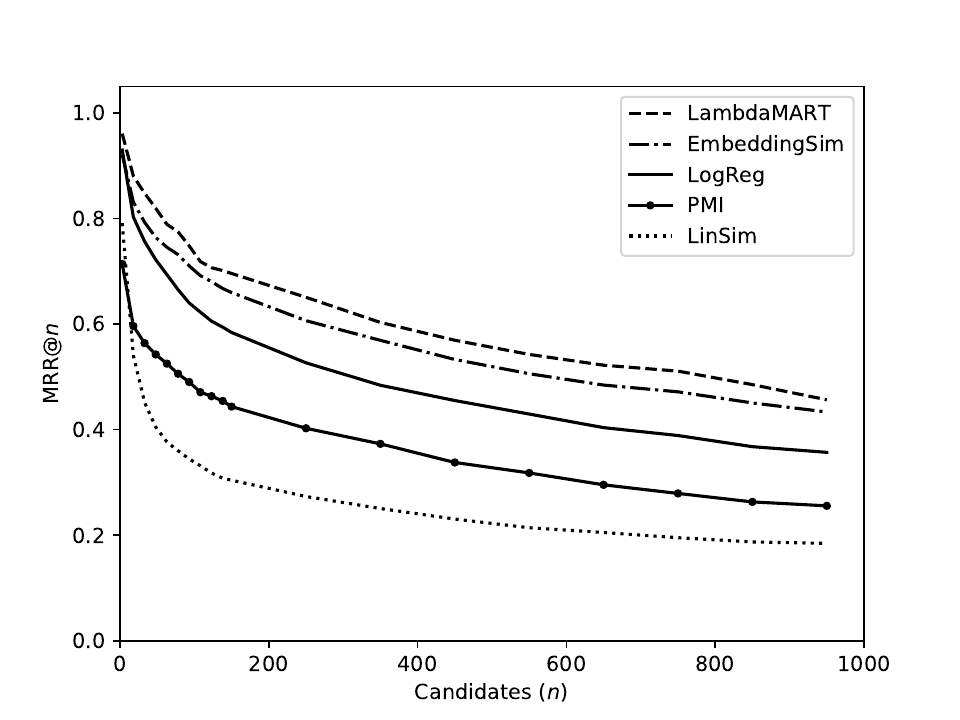}}
	\\
	\subfloat[Ranking evaluation (MAP)]{\label{subfig:map}
		\includegraphics[width=0.48\textwidth]{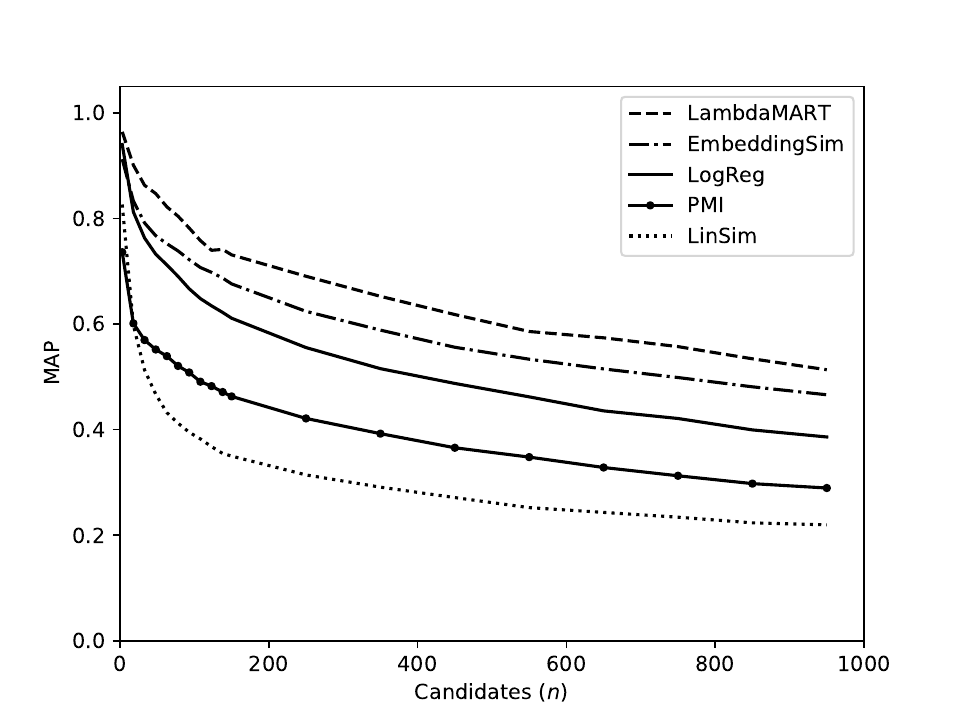}}
	\hspace{0.1cm}
	\subfloat[Ranking evaluation (Recall@$n$)]{\label{subfig:recall}
		\includegraphics[width=0.48\textwidth]{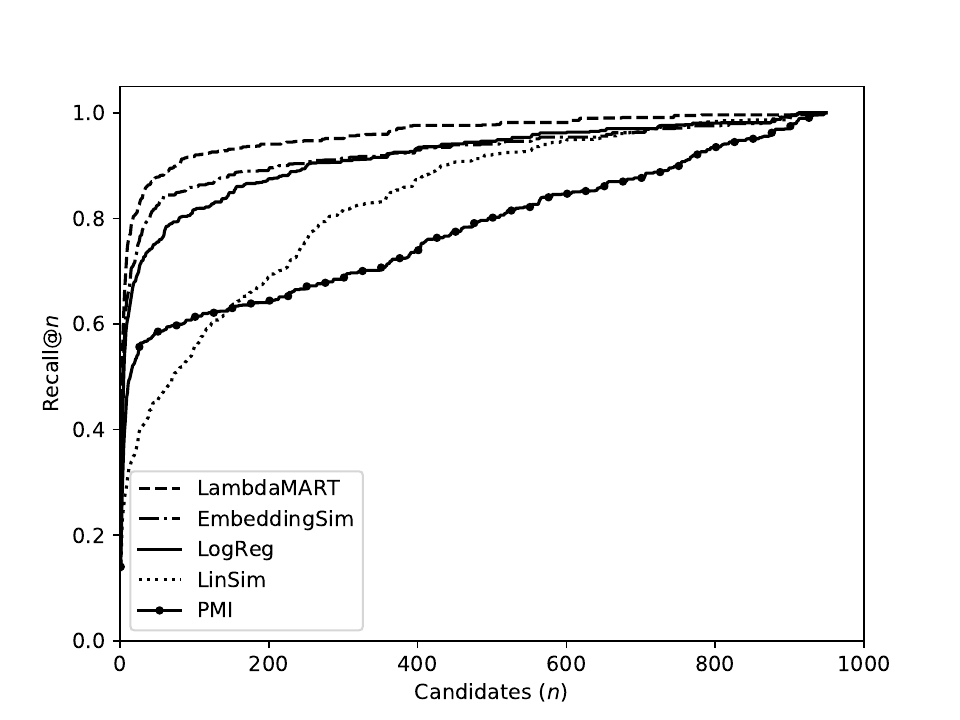}}
	\caption{Results on the TOEFL-style evaluation (top row) and relevance ranking evaluation (bottom row). While the top methods perform well in the TOEFL evaluation for low values of $n$, the performance of every method decreases as $n$ is increased to more realistic values. The EmbeddingSim baseline performs well on its own, but is surpassed by the LambdaMART model that incorporates it as a feature.}
	\label{fig:results} 
\end{figure}

\subsection{General TOEFL-Style Evaluation}\label{sec:ge}
As in the original work, we first perform a TOEFL-style evaluation to illustrate
the difficulty of the domain-specific synonym discovery task. In this evaluation,
methods are required to identify a target term's true synonym given one correct
synonym candidate and $n$ incorrect candidates. When $n=3$ this corresponds to
the TOEFL evaluation commonly used in prior work on discovering domain-independent
synonyms. The original work made the argument that this evaluation is
unrealistically easy and demonstrated that, in the consumer health domain,
methods are unable to accurately identify synonyms when $n$ is increased to
realistic values (e.g., $n=1000$). In this section we repeat the general
TOEFL-style evaluation in order to demonstrate that considering only $n=3$
incorrect candidates is still unrealistically easy with our Swedish corpus
focused on the building construction domain.

For each pair consisting of a target term and a correct synonym candidate,
we randomly sample $n$ incorrect candidates and feed the candidates as
input to a synonym discovery method. As in the original work, we aggregate
each method's predictions to calculate accuracy@$n$.
We additionally report MRR (Mean Reciprocal Rank), which is a more informative metric
because correct results in positions past rank 1 also contribute to the score.
The result are shown in Figure~\ref{subfig:acc} (accuracy) and Figure~\ref{subfig:mrr} (MRR).
While LogReg, LambdaMart, and EmbeddingSim perform well at low values of $n$,
their accuracy when approaching $n=1000$ is less than 50\%.
The methods perform similarly in terms of MRR.
While LogReg, the method from the original work, continues to outperform PMI and LinSim, the new EmbeddingSim
baseline performs substantially better.
This may be due to the fact that LogReg is a linear model and thus has difficulty weighting EmbeddingSim substantially
higher than its other features.
LambdaMART, the alternate learning to rank model evaluated in this work,
is able to outperform both LogReg and EmbeddingSim.
This illustrates that the synonym discovery task remains difficult
in the new domain. For use cases where recall is important, such as ours, the task is best approached as a ranking problem.

\subsection{Relevance Ranking Evaluation}\label{sec:re}
In this section we evaluate the synonym discovery methods' ability to rank
synonym candidates, so that they may be considered by a human editor.
For each target term, we feed every synonym candidate as input to a synonym
discovery method and calculate MAP (Mean Average Precision) and recall@$n$
over the resulting rankings. In the context of this task, recall@$n$ is
the more interpretable metric: it indicates the fraction of correct synonyms
that a human editor would find after reading through the top $n$ results.
The results are shown in Figure~\ref{subfig:map} (MAP) and Figure~\ref{subfig:recall} (recall@$n$).
In general the ranking of methods mirrors that from the TOEFL-style evaluation, with LambdaMART
performing best. 
The top three methods perform similarly for different values of $n$, whereas
PMI and LinSim perform differently. PMI performs better for low values of $n$,
while LinSim begins to outperform PMI at roughly $n=175$.
As in the previous evaluation, the new LambdaMART and EmbeddingSim methods outperform LogReg.
LambdaMART achieves 88\% recall at $n=50$, followed by EmbeddingSim with 82\% recall and
LogReg with 76\% recall. This illustrates that the top performing methods can produce a useful ranking
despite their low accuracy.

\begin{figure}[!ht]
  \centering
  \subfloat[MAP when single features are used.]{
  \includegraphics[width=0.47\textwidth]{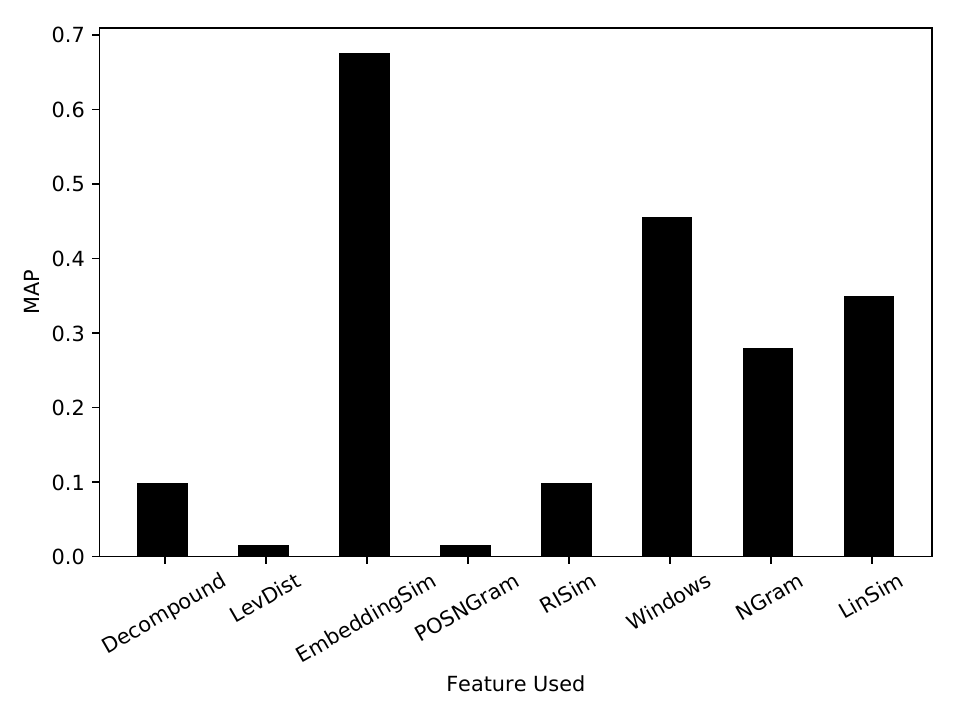}
  \label{fig:single}}
  \hspace{0.1cm}
  \subfloat[MAP (fraction of max) when single features are excluded from the model.]{
  \includegraphics[width=0.47\textwidth]{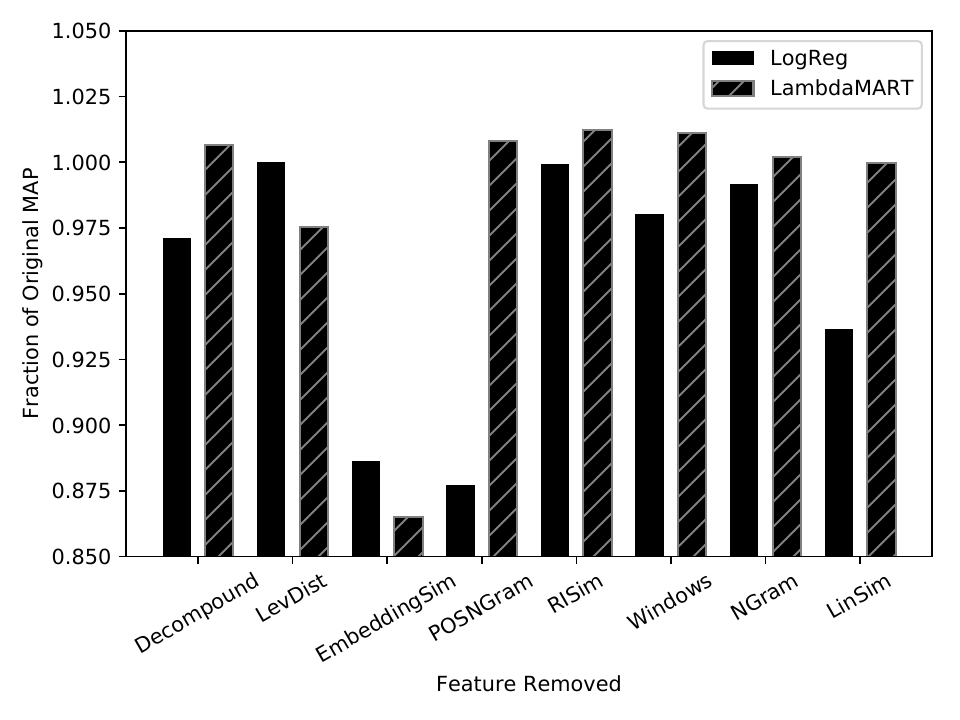}
  \label{fig:leftout}}
  \caption{MAP@150 when only one feature is used (left) and the fraction of the entire model's MAP@150
    when a single feature is excluded from the model (right). LevDist, the strongest feature in the original work,
    appears to have less utility in the new setting.
    Windows and LinSim, strong features in the original work,
    continue to perform well here.
  }
\end{figure}

\subsection{Feature Analysis}\label{sec:fa}
In this section we evaluate the contribution of individual features to the learning to rank models' performance.
The MAP@150 achieved by each individual feature is shown in Figure \ref{fig:single}.
EmbeddingSim performs substantially better than the other features.
Windows, NGram, and LinSim are the only other features to achieve a MAP above 0.3, with
Decompound, LevDist, and POSNGram performing poorly when used in isolation.
In the original work, LevDist was the best performing single feature, with Windows, RISim,
and the dependency context features (LinSim and Hagiwara) performing the next best.
In our new domain, Windows and LinSim continue to perform well, but LevDist and RISim perform poorly.
The difference in the domains and language may account for LevDist's decreased impact, since it is only
a useful feature when synonym candidates have significant character overlap with target terms.
EmbeddingSim was not included in the original work, but it is similar to RISim in that both methods
are intended to capture distributional semantics.

We analyze the decrease in each model's performance when a single feature is removed in Figure \ref{fig:leftout}.
The y-axis indicates each method's MAP as a fraction of the original MAP after a feature is removed.
For example, removing the EmbeddingSim feature reduces the performance of both LambdaMART and LogReg to
85-90\% of their MAPs when all features are used.
With the exception of LevDist and EmbeddingSim, the LambdaMART model consistently achieves a smaller decrease
in performance when any single feature is removed.
As in the single feature analysis, EmbeddingSim is the best performing feature, and RISim does not
contribute much to the models' performance.
While POSNGram performed poorly in isolation, removing the
feature decreases the performance of LogReg by approximately 12\%, indicating that it is providing a useful
signal used in conjunction with other features.
Similarly, removing Decompound or LevDist decreases LogReg's or LambdaMART's performance
by approximately 2.5\%, respectively, despite the fact that they performed poorly as single features.

\section{Generalizability}
In order to better understand the generalizability of the learning to rank approach to
synonym discovery, we discuss differences between this study and the original one
in terms of the methodology and results.

\paraHdSpace{Corpus Creation}
Due to its focus on identifying synonyms of medical side effects, the original study used a corpus of
400,000 English forum posts related to health. In this study we focused on Swedish language synonyms in
the building construction domain. This is a formal, specialized domain in comparison to health-related social media,
which made it more difficult to identify documents containing target terms or synonym candidates.
To create a corpus with sufficient term co-occurrence information, we created a Swedish language corpus
that was both larger and less homogeneous than the corpus used in the original study: 4.2 million Webpages
from Wikipedia, the Swedish Transportation Agency (Trafikverket), and searches against the Bing API.

\paraHd{Preprocessing}
While the preprocessing details differed in this study due to the change in languages, the techniques
used were conceptually similar to those used in the original study. We use MaltParser in place of RASP3,
and efselab's tokenization and lemmatizer in place of NLTK and the Porter stemmer.

\paraHd{Features and Method}
We introduced two new features that were not present in the original study.
Motivated by prior work that showed decompounding can improve recall in the German language \cite{Braschler2004},
we introduced the Decompounder feature. This feature uses the SECOS decompounder \cite{riedl2016secos}
to split compound nouns into their components and measures the overlap between a target term's and candidate term's
components. Figure \ref{fig:leftout} suggests that this feature slightly contributes to the performance of the LogReg
method, but does not positively influence the performance of LambdaMART.
EmbeddingSim, which computes the similarity between FastText embeddings, is the second new feature we introduced.
FastText considers character n-grams when representing terms, which may make the Decompound feature redundant.
We additionally introduced experiments on a new learning to rank model, LambdaMART, in order to compare its
performance with the LogReg model used in the original work. We found that LambdaMART substantially outperformed
LogReg, indicating the utility of using a more advanced model.
We found that methods generally performed better on our domain and corpus. For example, LogReg achieved
50\% recall@50 in the original study, whereas it achieves 76\% recall@50 in this work. It is difficult
to attribute these performance differences to specific factors, with the language, domain, and language register
(i.e., professional language in this study and casual, lay language in the original study) all differing
between the two studies.


\section{Conclusions}
In this work we replicated the synonym discovery method proposed by Yates et al.~\cite{Yates14} in a
new language (i.e., Swedish rather than English) and in a new domain (i.e., building construction rather
than medical side effects). We found that in the new domain, the proposed LogReg method outperformed
the PMI baseline as before. Motivated by methodological advances and the difference in languages, we introduced
two new features and an alternate learning to rank method which we found to outperform the original approach.

These results provide evidence that \textit{(1)} synonym discovery can be effectively approached as a learning to rank problem and \textit{(2)} the features proposed in the original work are robust to changes in both domain and language. While our replication cannot provide evidence that the new EmbeddingSim feature works well in other settings, it does provide evidence that EmbeddingSim does not make the features used in the original work obsolete.

%
%
%
\bibliographystyle{splncs04}
\bibliography{references}
\end{document}